\documentclass[doublecol]{epl2} 
\usepackage{amsmath}
\usepackage{epsfig}
\usepackage{dcolumn}
\usepackage{bm}
\usepackage{amsfonts}
\usepackage{graphicx}
\usepackage{color}

\title{Current noise spectrum in a solvable model of tunneling Fermi-edge singularity.}
\author{V.V. Ponomarenko\inst{1} and I. A. Larkin\inst{2}}
\institute{\inst{1}Nonlinearity and Complexity Research Group, Aston University, Birmingham
B4 7ET, United Kingdom\\
 \inst{2}Institute of
Microelectronics Technology RAS, 142432 Chernogolovka, Russia}

\abstract{
We consider tunneling of spinless electrons from a single channel
emitter into an empty collector through an interacting resonant
level of the quantum dot (QD). When all Coulomb screening of sudden
charge variations of the dot during the tunneling is realized by
the emitter channel, the system is mapped onto an exactly
solvable model of a dissipative qubit.
The qubit density matrix evolution is described with a generalized  Bloch
equation which permits us to count the tunneling electrons and
find the charge transfer statistics.
The two generating functions of the counting statistics of the charge transferred
during the QD evolutions form
its stationary and empty state
have been expressed through each other.
It is used to calculate the spectrum of the steady current noise and
demonstrate occurrence of the bifurcation of its single zero-frequency minimum into
two finite-frequency dips due to the qubit coherent dynamics.
}

\begin{document}

\maketitle

The Fermi-edge singularity (FES) resulting \cite{1,2} from
the reconstruction of the Fermi sea
of conduction electrons under a sudden change of a local potential
have been primarily
observed \cite{3,4} as a power-law singularity in X-ray absorption
spectra.
A similar phenomenon of the FES in
transport of spinless electrons through a quantum dot (QD) was
predicted \cite{5}  in the perturbative regime when a localized QD level is
below the Fermi level of the emitter in its proximity and the
collector is effectively empty (or in equivalent formulation through
the particle-hole symmetry) and the tunneling rate of the emitter is
sufficiently small.
Then,  the subsequent
separated in time electron tunnelings from the emitter vary the localized level
charge and generate sudden changes of the scattering potential
leading to the FES in the I-V curves at the voltage threshold
corresponding to the resonance.
Direct observation of these
perturbative results in experiments \cite{geim,6,7,8,9,lar}, however,
is complicated due to the uncontrolled effects such as
of the finite life time of electrons (level broadering of the localized state of the
QD), temperature smearing and variation of tunneling parameters due to application
of the bias voltage.
Therefore, it has been suggested \cite{epl} that the true FES nature of
a threshold peak in the I-V dependence
can be verified through observation of
the oscillatory behavior of a corresponding time-dependent transient current.
Indeed, in the FES theory \cite{1,2}  appearance of such a threshold peak
signals formation
of a two-level system of the exciton electron-hole pair or qubit
in the tunneling channel at the QD. The qubit undergoes dissipative dynamics
characterized \cite{L,Sch}, in the absence of the collector tunneling, by the  oscillations
of the levels occupation.
By studying an exactly solvable model of the FES we earlier demonstrated
that for a wide range of the model parameters the qubit dynamics also
manifest themselves through the resonant features of the {\it a.c.} response \cite{prb}
and in the oscillating behavior \cite{epl} of the collector
transient current, in particular, when the QD evolves from its empty state.
Although a possible observation of these oscillations would give the most
direct verification of the nature of the I-V threshold peaks,
in the recent experiments \cite{8,noise} the low-temperature noise measurements
have been considered for this purpose.

For this reason, in this work we study quantum fluctuations of the steady tunneling current
into the collector in the exactly solvable model of the FES.
This model describes a simplified, but still realistic system
of spinless electrons tunneling from a single channel
emitter into an empty collector through an interacting resonant
level of QD, when all Coulomb screening of sudden
variations of the charge in the QD is realized by
the emitter channel. In order to see traces of
the qubit dynamics in the shot noise spectrum of the
current we apply methods of the full counting statistics by describing
the qubit density matrix evolution with a generalized  Bloch
equation, which permits us to count the tunneling electrons and
find the charge transfer statistics.
The two generating functions of statistics of the charge transferred during the QD evolution
from its stationary and empty states
are expressed through each other. This relation
is further used to establish the direct connection between the spectrum of the steady
current shot noise and Fourier transformation of the time-dependent transient current
produced in the process of the QD empty state evolution.
The final expression for  the current noise spectrum is analyzed to
conclude how the spectral features reflect the oscillating behavior of
the time-dependent transient current.

\section{Model} The system we consider below is described with Hamiltonian
$\mathcal{H}=\mathcal{H}_{res}+\mathcal{H}_C$ consisting of the
one-particle Hamiltonian of resonant tunneling of spinless
electrons and the Coulomb interaction between instant charge
variations of the dot and electrons in the emitter. The resonant
tunneling Hamiltonian takes the following form
\begin{equation}
\mathcal{H}_{res}=-\epsilon _{d} d^+d+\sum_{a=e,c}\mathcal{H}_0[\psi_a]+
w_a(d^+\psi_a(0)+h.c.) \ ,  \label{hres}
\end{equation}
where the first term represents the resonant level of the dot, whose energy
is $-\epsilon _{d}$. Electrons in the emitter (collector) are described with the chiral
Fermi fields $\psi_a(x),a=e(c)$, whose dynamics is governed by the
Hamiltonian $\mathcal{H}_0[\psi]==-i\!\! \int\! dx \psi^+(x) \partial_x
\psi(x) \ ( \hbar=1) $ with the Fermi level equal to zero or drawn to $%
-\infty$, respectively, and $w_a$ are the correspondent tunneling
amplitudes. The Coulomb interaction in the Hamiltonian $\mathcal{H}$ is
introduced as
\begin{equation}
\mathcal{H}_C=U_C \psi_e^+(0)\psi_e(0)(d^+d-1/2) \ .  \label{hc}
\end{equation}
Its strength parameter $U_C$ defines the scattering phase
variation $\delta$ for electrons in the emitter channel
and therefore the change of the localized charge in the emitter
$\Delta n=\delta/\pi \ \ (e=1)$, which we assume provides the
perfect screening of the QD charge: $\Delta n=-1$.

After implementation of  bosonization of the emitter Fermi field $%
\psi_e(x)=\sqrt{\frac{D}{2\pi} }\eta e^{i\phi(x)}$,
where $\eta$ denotes an
auxiliary Majorana fermion, $D$ is the large Fermi energy of the emitter,
and the chiral Bose field $\phi(x)$ satisfies $[\partial_x\phi(x),\phi(y)]=i2\pi%
\delta(x-y) $,
and further completion of a standard rotation \cite{schotte},  under
the above screening assumption
we have transformed \cite{epl} $\mathcal{H}$ into the
Hamiltonian of the dissipative two-level system or qubit:
\begin{eqnarray}
\mathcal{H}_{Q}=-\epsilon _{d} d^+d+\mathcal{H}_0[\psi_c]+ w_c(\psi^+_c(0)d+h.c.)
\notag \\
+\Delta \eta (d- d^+) \ ,  \label{hq}
\end{eqnarray}
where $\Delta = \sqrt{\frac{D}{2\pi }}w_{e}$ and
the time
dependent correlator of electrons in the empty collector $<\psi_c(t)\psi^+_c(0)>=%
\delta(t)$ has been used to drop
the bosonic exponents in the third term on the right-hand side in (\ref{hq}).

\section{Charge counting Bloch equation for the qubit evolution}
We use this Hamiltonian to describe the dissipative evolution of the qubit
density matrix $\rho _{a,b}(t)$, where $a,b=0,1$ denote the empty and filled
levels, respectively. In the absence of the tunneling into the collector at $%
w_{c}=0$, $\mathcal{H}_{Q}$ in Eq. (\ref{hq}) transforms through the
substitutions of $\eta (d-d^{+})=\sigma _{1}$ and $d^{+}d=(1-\sigma _{3})/2$
( $\sigma _{1,3}$ are the corresponding Pauli matrices) into the Hamiltonian
$\mathcal{H}_{S}$ of a spin $1/2$ rotating in the magnetic field
$\mathbf{h}=(2\Delta , 0,\epsilon_{d})^{T}$ with the frequency
$\omega_0=\sqrt{4\Delta ^{2}+\epsilon _{d}^{2}}$ .
Then the evolution equation follows from
\begin{equation}
\partial _{t}\rho (t)=i[\rho (t),\mathcal{H}_{S}]\ .  \label{rhos}
\end{equation}%
To incorporate in it the dissipation effect due to tunneling into the empty
collector we apply the diagrammatic perturbative expansion of the S-matrix
defined by the Hamiltonian (\ref{hq}) in the tunneling amplitudes $w_{e,c}$
in the Keldysh technique. This permits us to integrate out the collector
Fermi field in the following way. At an arbitrary time $t$ each diagram
ascribes indexes $a(t_{+})$ and $b(t_{-})$ of the qubit states to the upper
and lower branches of the time-loop Keldysh contour. This corresponds to the
qubit state characterized by the $\rho _{a,b}(t)$ element of the density
matrix. The expansion in $w_{e}$ produces two-leg vertices in each line,
which change the line index into the opposite one. Their effect on the
density matrix evolution has been already included in Eq. (\ref{rhos}).
In addition, each line with index $1$ acquires two-leg diagonal vertices
produced by the electronic correlators $<\psi _{c}(t_{\alpha })\psi
_{c}^{+}(t_{\alpha }^{\prime }) >,\ \alpha =\pm $. They result in the
additional contributions to the density matrix variation: $\Delta \partial
_{t}\rho _{10}(t)=-\Gamma \rho _{10}(t),\ \Delta \partial _{t}\rho
_{01}(t)=-\Gamma \rho _{01}(t),\ \Delta \partial _{t}\rho _{11}(t)=-2\Gamma
\rho _{11}(t),\ \Gamma =w_{c}^{2}/2$. Next,
to count the electron tunnelings into the collector we ascribe \cite{lll} the opposite
phases to the collector tunneling amplitude $w_{c} \exp\{\pm i \chi/2\}$ along the
upper and lower Keldysh contour branch, correspondingly. These phases
do not affect the above contributions, which do not mix the amplitudes of
the different branches.
Then there are also vertical fermion
lines from the upper branch to the lower one due to the non-vanishing
correlator $<\psi _{c}(t_{-})\psi _{c}^{+}(t_{+}^{\prime })>$, which lead to
the variation affected by the phase difference as follows
$\Delta \partial _{t}\rho _{00}(t)=2\Gamma w \rho _{11}(t), \ w=\exp\{i\chi\}$.
Incorporating these additional terms into Eq. (\ref{rhos}) we come to
the modified quantum master equation
\begin{eqnarray}
\partial _{t}\rho (t,w)=i[\rho,\mathcal{H}_{S}]-\Gamma|1><1|\rho -\Gamma\rho|1><1| \notag \\
+2w\Gamma|0><1|\rho|1><0| .  \label{lindblad}
\end{eqnarray}
for the qubit density matrix evolution and also counting the charge transfer. Here the vectors
$|0>=(1,0)^T$ and $|1>=(0.1)^T$ describe the empty and filled QD, respectively. This equation is of
a type  known in the theory of open quantum systems as the GKSL equations \cite{GKS,Lindblad} and is exact
in our case.
Solving Eq. (\ref{lindblad}) with some initial $\rho(0)$ independent of $w$ at $t=0$, we find
the generating function $P(w,t)$ of the full counting statistics of the charge transfer
by calculating the trace of the density matrix:
$P(w,t)=Tr[\rho(w,t)]=\sum_{n=0}^\infty P_n(t)w^n$.

Making use of the four-component Bloch vector  $\mathbf{a}(t,w)$
we represent the trace non-conserving density matrix as
$\rho (t,w)=[a_0(t,w)+\sum_{l}a_{l}(t)\sigma_{l}]/2 $, where  the additional
component $a_0=P(w,t)$ evolves from its initial value $a_0(0)=1$ and stays equal to one
at $w=1$, but as a function of $w$ it gives us the generating function of
charge transfer during the process time $t$.
Substitution of this density matrix representation into Eq. (\ref{lindblad}) results in
the following  evolution equation for the Bloch vector $\mathbf{a}(t,w)$
\begin{equation}
\partial _{t}\mathbf{a}(t,w)=M(w)\cdot \mathbf{a}(t,w)\ ,  \label{dadt}
\end{equation}%
where $M(w)$ stands for the matrix:
\begin{equation}
M\!=\!\left(
\begin{array}{llll}
0 & 0 & 0 & 0 \\
0 & -\Gamma & -\epsilon _{D} & 0 \\
0 & \epsilon _{D} & -\Gamma & -2\Delta \\
2 \Gamma & 0 & 2\Delta & -2\Gamma%
\end{array}%
\right)
 + (w-1) \Gamma |\mathbf{e}_E\!><\mathbf{e}_F|  \label{M}
\end{equation}%
and the ket and bra vectors $|\mathbf{e_E}>=(1,0,0,1)^T, \ <\mathbf{e_F}|=(1,0,0,-1)$
define the empty and filled QD state, respectively.

The general solution to Eq. (\ref{dadt}) describing the evolution of
the Bloch vector starting from its value $\mathbf{a}(0)$ independent of $w$ at zero time can be
found through the Laplace transformation in the following form:
\begin{equation}
\mathbf{a}(t,w)=\int_{C}\frac{dz e^{zt}}{2 \pi i}\left[ z-M(w)\right]^{-1} \mathbf{a}(0) \ ,
\label{atw}
\end{equation}%
where the integration contour $C$ coincides with the imaginary axis shifting
to the right far enough to have all poles of the integral on its left side.
Writing the inverse matrix in the standard form
$[z-M(w)]^{-1}= [z-M(w)]_A/\det[z-M(w)]$, where $[z-M(w)]_A$ denotes the corresponding matrix
of the algebraic complements, we conclude that these poles are equal
to four roots of its determinant $\det \left[ z-M(w)\right] \equiv p_4(z+\Gamma)$,
which is
\begin{equation}
p_4(x)=x^{4}+(4\Delta ^{2}+\epsilon _{d}^{2}-\Gamma^2) x^{2}-4\Delta ^{2}\Gamma w x -\Gamma^2
\epsilon_{d}^{2} \ .  \label{p4}
\end{equation}
Then the general form of the generating function follows from (\ref{atw}) as
\begin{equation}
P(t,w)=\int_{C}\frac{dz e^{zt}}{2\pi i}\frac{g_a(z+\Gamma,w)}{p_4(z+\Gamma)} \ ,
\label{ptw}
\end{equation}%
where $g_a(z+\Gamma,w) \equiv <\mathbf{e0}|[z-M(w)]_A|\mathbf{a}(0)>$ contrary to $p_4(x)$
depends also on the initial Bloch vector and $<\mathbf{e0}|=(1,0,0,0) $.
We are interested to consider the process starting from the stationary Bloch
vector $\mathbf{a}(0)=\mathbf{a}^{st}$ defined by $M(1)\mathbf{a}^{st}=0$.
As we show below this process, in fact,
is determined by the Bloch vector evolution starting from the empty QD.

Solving  $M(1)\mathbf{a}^{st}=0$ with $M(1)$ from Eq. and $a^{st}_0=1$ we find
the stationary  Bloch vector $\mathbf{a}^{st}=[1,\mathbf{a}^T_\infty ]^T$, where:
\begin{equation}
\mathbf{a}_\infty=\frac{[2\epsilon _{d}\Delta
,-2\Delta \Gamma ,(\epsilon _{d}^{2}+\Gamma ^{2})]^{T}}{\left( \epsilon
_{d}^{2}+\Gamma ^{2}+2\Delta ^{2}\right) }\ .  \label{ainfty}
\end{equation}%
In general, an instant tunneling current $I(t)$ into the empty collector
directly measures the diagonal matrix element of the qubit density matrix
\cite{us} through their relation
\begin{equation}
I(t)=2\Gamma \rho _{11}(t,1)=\Gamma \lbrack 1-a_{3}(t,1)] \, . \label{I-t}
\end{equation}%
It gives us the stationary tunneling current as $I_{0}=2\Gamma
\Delta^{2}/(2\Delta ^{2}+\Gamma ^{2}+\epsilon _{d}^{2})$. At $\Gamma \gg
\Delta $ this expression coincides with the perturbative results of
\cite{5,lar}.

Substitution of $M(w)$ from Eq. (\ref{M}) into the denominator of the integrand
on the right side of Eq, (\ref{atw}) and further its expansion in $(w-1)$ brings
up the following expression for the Bloch vector evolution:
\begin{eqnarray}
\mathbf{a}(t,w)\!\!\!\!&=&\!\!\!\!\int_{C}\frac{dz }{2\pi i}\frac{ e^{zt}}{[z-M(1)]}   \label{atw2} \\
&\times&\!\!\!\! \sum_{n=0} \left[(w-1) \Gamma |\mathbf{e}_E>
<\mathbf{e}_F|\left(z-M(1)\right)^{-1}\right]^n\!\mathbf{a}(0) \ .\nonumber
\end{eqnarray}%
For the initial vector  $\mathbf{a}(0)=\mathbf{a}^{st}$ this expression transforms into
\begin{eqnarray}
\mathbf{a}^{st}(t,w)=\mathbf{a}^{st}(0)+(w-1) I_0
\int_{C}\frac{dz }{2\pi i z}\frac{e^{zt}}{[z-M(1)]}\,
\mathbf{e}_E \notag \\
\times \sum_{n=0} \left(<\mathbf{e}_F|\frac{(w-1) \Gamma}
{z-M(1)}|\mathbf{e}_E>\right)^n  \
\label{ast1}
\end{eqnarray}%
due to the properties of the stationary Bloch vector discussed above.
On the other hand, for the evolution from the empty QD and the choice  $\mathbf{a}(0)=\mathbf{e}_E$
Eq. (\ref{atw2}) can be re-written as
\begin{eqnarray}
\mathbf{a}^{E}(t,w)\!=\!\!
\int_{C}\frac{dz }{2\pi i}\!\!\!\!&&\!\!\!\!\!\frac{e^{zt}}{[z-M(1)]}\,
\mathbf{e}_E   \label{aE} \\
& & \times\sum_{n=0} \left(<\mathbf{e}_F|\frac{(w-1) \Gamma}
{z-M(1)}|\mathbf{e}_E\!>\right)^n \!\!\! . \notag
\end{eqnarray}%
From comparison of these two expressions we find the relation between
the two Bloch vectors:
\begin{equation}
\mathbf{a}^{st}(t,w)=\mathbf{a}^{st}(0)+(w-1) I_0
\int_0^t d\tau \, \mathbf{a}^{E}(\tau,w)\, .
\label{ast2}
\end{equation}
The relation (\ref{ast2}) between the zero components of the Bloch vectors shows that
the generating function $P^{st}(t,w)$ of the charge transfer statistics in
the process starting with QD in the stationary state can be found from the generating function
for the process starting from the empty QD. By differentiating it with respect of the time one
can rewrite this relation as
\begin{equation}
\partial_t P^{st}(t,w)=(w-1) I_0 P(t,w) \theta(t)\ ,
\label{Pst}
\end{equation}%
where the Heavyside step function $\theta(t)$ starts counting the charge transfer at $t=0$.
It is straightforward to see from Eq. (\ref{Pst}) that in the steady process
$<I>_{st}=\partial_w \partial_t P^{st}(t,1)=I_0$ and similarly one can relate higher
current correlators.
Therefore, it suffices below to focus our study on the generating function
$P(t,w)$ for the process starting from the empty QD.
In this case  $g_E(z+\Gamma,w) \equiv <\mathbf{e0}|[z-M(w)]_A|\mathbf{e}_E>$ is calculated as:
\begin{equation}
g_E(x)=x^{3} +\Gamma x^{2}+(4\Delta ^{2}+\epsilon _{d}^{2})x+\Gamma \epsilon_{d}^{2} \ ,
\label{fE}
\end{equation}
which does not depend on $w$.
\begin{figure}
\onefigure{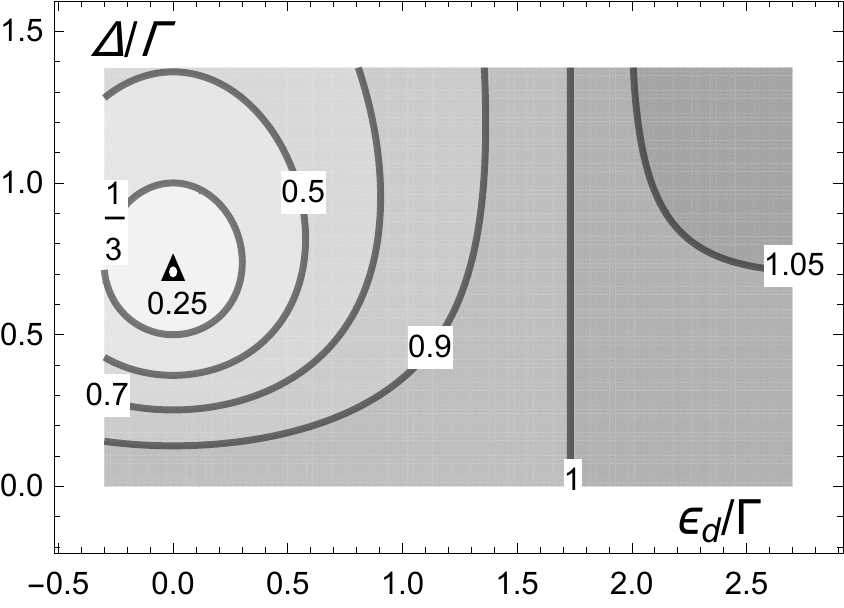} \caption{Contour plot of the
Fano factor $F_2(0)$ in Eq. (\ref{f2})
 as a function of the $\epsilon_{d}/\Gamma$ and $\Delta/\Gamma$.
 White point in the black triangle corresponds to the absolute minimum of $F_2(0)$.
}
\label{fig:I01}
\end{figure}

\begin{figure}
\onefigure{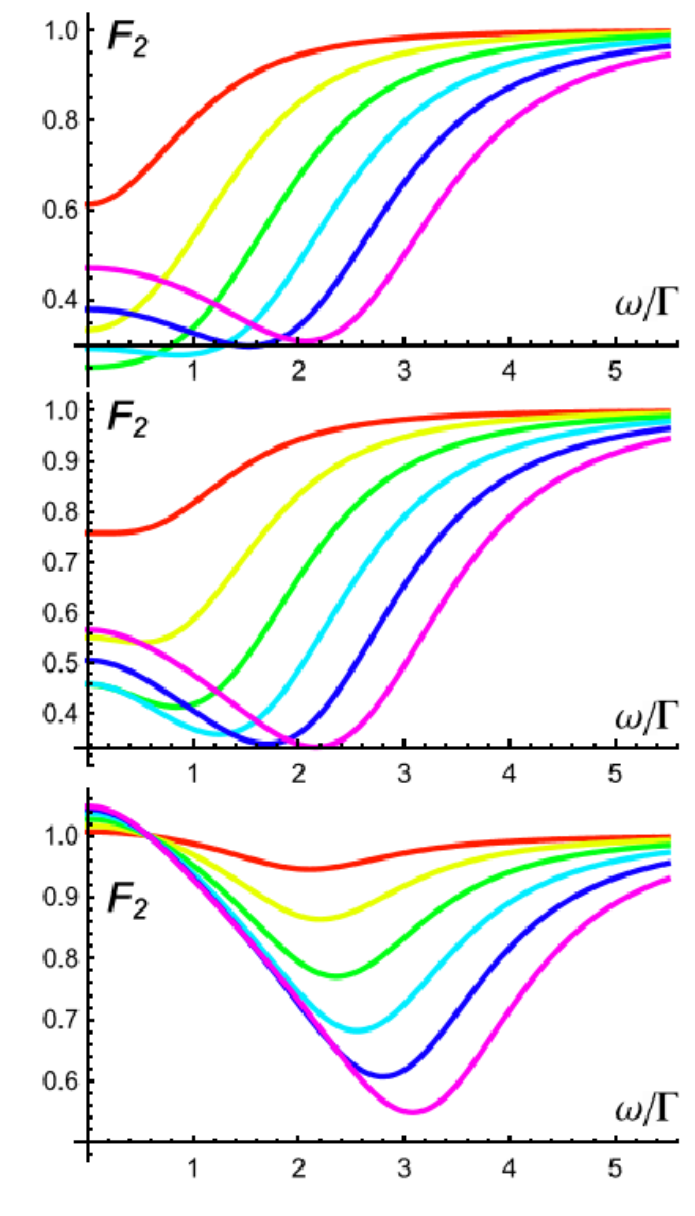}
 \caption{Plot of the
Fano factor $F_2(\omega)$ in Eq. \ref{f2omega} as a function of $\omega$.
Red, yellow, green, light blue, blue and purple lines correspond to
the parameter $\Delta=$ 0.3, 0.5, 0.7, 0.9,1.1 and 1.3.
Upper panel corresponds to $\epsilon_{d}=$ 0,
medium panel corresponds to $\epsilon_{d}=$ 0.5 and
low  panel corresponds to $\epsilon_{d}=$ 2.
}
\label{fig:I02}
\end{figure}
\begin{figure}
\onefigure{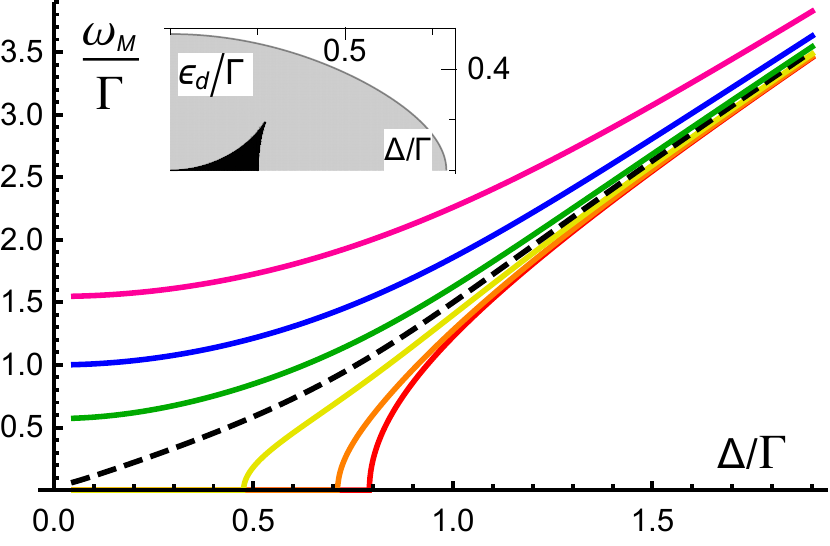} \caption{Plot of the minimum location $\omega_M$ of
the Fano factor $F_2(\omega)$ in Eq. \ref{f2omega} as a function of $\Delta$.
Red, orange, yellow, green,  blue and purple lines correspond to
the parameter $\epsilon_{d}/\Gamma$=
 0, 0.2, 0.4, 0.7, 1, and 1.5. Black dashed curve corresponds to the bifurcation point
$\epsilon_{d}/\Gamma \approx 0.54$.
Inset: grey area corresponds to the single Fano factor minimum at $\omega_M=0$,
black area from \cite{epl} shows where the transient current does not oscillate.}
\label{fig:I03}
\end{figure}

\section{Spectrum of the current noise}

In order to calculate the current noise spectrum $S(\omega)$
defined as the Fourier transformation
of the real part of the current-current correlator
\begin{eqnarray}
S(\omega)=\frac{1}{2}\int_{-\infty}^{\infty }dt \,e^{i\omega t}<\{I(t),I(0)\}_+> \notag \\
= Re[\int_{0}^{\infty }dt \, e^{i\omega t}<\{I(t),I(0)\}_+>]  \label{S}
\end{eqnarray}
we need to express the time-dependent correlator
$<\{I(t),I(0)\}_+>=Sp(\{I(t),I(0)\}_+\rho_{st}(0))$
in terms of the generating function $P(t,w)$.
This can be done through the following relation:
\begin{equation}
\partial_t<N^2(t)>_{st}=\int_{0}^{t }dt'<\{I(t'),I(0)\}_+> \, .
\label{N2st} \\
\end{equation}
Since its left-hand side is equal to $\partial_t (w\partial_w)^2 P^{st}(t,w)$ at $w=1$,
we obtain the expression in question making use of Eq. (\ref{Pst}) in the following form:
\begin{equation}
<\{I(t),I_{0}\}_+>= I_0\delta(t)+2I_0 \partial_w\partial_t P(t,w)|_{w=1} \ .
\label{II+} \\
\end{equation}
The derivative of the generating function on the right-hand side of Eq. (\ref{II+})
coincides with the transient
current $<I(t)>_E$ in the tunneling  process, which starts
from the empty QD state. Its oscillating behavior has been suggested \cite{epl} as an
observable manifestation of the qubit dynamics at the FES.
Therefore the current noise spectrum relates to the spectral
decomposition of this transient current as follows
\begin{equation}
S(\omega)=I_0+2I_0
\int_{0}^{\infty }\!\!dt \cos(\omega t) <I(t)>_E \,  \label{SIE}
\end{equation}
and should reflect its oscillatory features.

Substituting here the  $P(t,w)$ derivative expression through the inverse
Laplace transformation in Eq. (\ref{ptw}) and taking the time integral and then
the contour integral after closing
the contour in the right half-plane we come to
\begin{equation}
S(\omega)=I_{0}+2I_{0}Re[-i\omega g_E(\Gamma-i\omega)
\partial_w p_4^{-1}(\Gamma-i\omega,w)]|_{w=1} \, ,\  \label{S1} \
\end{equation}
where the functions $P_4$ and $g_E$ are specified in Eqs. (\ref{p4},\ref{fE}).
Making use of their explicit expressions we calculate the right-hand side in Eq. (\ref{S1})
and write the final result in the normalized form $S(\omega)/I_0 \equiv F_2(\omega)$
of the frequency-dependent Fano factor
\begin{eqnarray}
\!\!\!\!\!\!\!&&\!\!\!\!\!F_2(\omega)=1 - \label{f2omega} \\
\!\!\!\!\!\!\!\!&&\!\!\!\!\!\frac{8\Gamma^2 \Delta^2 (3(\Gamma^2+\omega^2)-\epsilon^2_{d})}
{4\Gamma^2 (\Gamma^2\!+\epsilon^2_{d}\!+2\Delta^2\!-2\omega^2)^2\! +\omega^2
(\omega^2\!-\epsilon^2_d-5\Gamma^2\!-4\Delta^2)^2} \, . \notag
\end{eqnarray}
Its zero-frequency limit reduces to
\begin{equation}
F_2(0)=1+\frac{2\Delta ^{2}(\epsilon _{d}^{2}-3\Gamma^2)}{(\epsilon_{d}^{2}+\Gamma^2 +2\Delta ^{2})^2}
 \  \label{f2}
\end{equation}
and shows in Fig.\ref{fig:I01} the clear border given by $\epsilon _{d}^{2}=3\Gamma^2 $ between the
sub-Poissonian distributions of the current fluctuations near the resonance at $\epsilon_{d}=0$
and the super-Poissonian ones far from it. The Fano factor $F_2(0)$ takes its smallest values
at the resonance, where it reaches its minimum $F_2=1/4$ at $\Delta=\Gamma/\sqrt{2}$.
The frequency dependence of $F_2(\omega)$ at the resonance
follows from Eq. (\ref{f2omega}) as:
\begin{equation}
F_2=1-\frac{24\Delta ^{2}\Gamma^2}{4\Gamma^4+(16\Delta ^{2}+5\omega^2)\Gamma^2 +(\omega^2-4\Delta ^{2})^2}
 \,  \label{f2resonance}
\end{equation}
and is depicted in the upper panel of Fig.\ref{fig:I02}.
Its single zero-frequency minimum at small $\Delta$ splits with increase of $\Delta$ into two minima located
at the finite frequencies $\pm \omega_M, \, \omega_M=2\sqrt{\Delta ^{2}-5 \Gamma^2/8}  $,
when $\Delta \ge \sqrt{5 /8} \Gamma \approx 0.79 \Gamma $, which are equal to
\begin{equation}
\min_\omega F_2(\omega)=\frac{12\Delta ^{2}-9\Gamma^2/4}{36\Delta ^{2}-9\Gamma^2/4 } \ \le \frac{1}{3}
 \,  . \label{minf2}
\end{equation}
This two dips split in the frequency dependence of $F_2(\omega)$ signals occurrence of the oscillations in the
time-dependent transient current \cite{epl}, but with the higher frequency \cite{prb}
$\omega_I=2\sqrt{\Delta ^{2}-\Gamma^2/16}$ than $\omega_M$.

Moving $\epsilon _{d}$ out of the resonance one finds the increase of $\omega_M$ and
that more minimum positions at smaller $\Delta$ split and shift away from the zero frequency
as illustrated by the medium and low panels of Figs.\ref{fig:I02},\ref{fig:I03}.
Minimization of the right-hand side of  Eq. (\ref{f2omega}) defines
the parametric region of $\omega_M=0$ with the inequality:
\begin{eqnarray}
F(\epsilon_d,\Delta,\Gamma) \equiv \epsilon _d^6+
\left(51
   \Gamma ^4+32 \Gamma ^2 \Delta ^2+16 \Delta
   ^4\right) \epsilon _d^2 \nonumber \\
+\left(3 \Gamma
   ^2+8 \Delta ^2\right) \epsilon _d^4
+24 \Gamma ^4 \Delta ^2 - 15 \Gamma ^6\le 0
 \ .
 \label{area}
\end{eqnarray}
It is depicted in the inset in Fig. \ref{fig:I03} as the grey area that covers the black
one corresponding \cite{epl} to the non-oscillating transient current behavior. The equality
in Eq. (\ref{area}) defines the grey area boundary serving as a bifurcation line,
on crossing of which outward the single zero-frequency minimum of $F_2(\omega)$ splits
into the two dips. These dips are located at $\pm \omega_M$, where $\omega_M^2$ coincides
with the real positive root of the cubic equation:
\begin{eqnarray}
6X^3+3\left(9 \Gamma^2-3 \epsilon _d^2 -8 \Delta ^2\right) X^2+4
\left(3 \Gamma^2-\epsilon _d^2 \right)  \\
\times \left(3  \Gamma ^2 -  \epsilon _d^2- 4 \Delta ^2\right) X - F(\epsilon_d,\Delta,\Gamma)
=0 \ .\nonumber
 \label{omegam}
\end{eqnarray}
Near the bifurcation line $\omega^2_M$ is small and reduces to
\begin{equation}
\omega^2_M=\frac{\theta(F) F(\epsilon_d,\Delta,\Gamma)}{4 \left(3 \Gamma^2-\epsilon _d^2 \right)
\left(3  \Gamma ^2 -  \omega^2_0 \right)}
 \,  , \label{omegaMnear}
\end{equation}
whereas far from the bifurcation line it is given asymptotically in small $\Gamma/\omega_0$ by
\begin{equation}
\omega^2_M=\omega^2_0- 10 \Gamma^2
\frac{\Delta^2}{\omega^2_0}+15 \Gamma^4 \frac{\omega^4_0+6\omega^2_0 \Delta^2-40 \Delta^4 }{4 \omega^6_0}
 \,  . \label{omegaMfar}
\end{equation}
Note in this limit $\omega^2_M$ can be larger than $\omega^2_0$ if $\epsilon _d^2 \gg \Gamma^2 >\Delta^2 $.
Substitution of the asymptotics (\ref{omegaMfar})
into Eq. (\ref{f2omega}) gives us the Fano factor asymptotics as
\begin{equation}
F_2 (\omega_M)=\frac{2 \Delta ^2+\epsilon_d^2}{6 \Delta ^2+\epsilon_d^2}
+O\left(\frac{\Gamma ^2 }
{\omega_0^2}\right) \,  . \label{minMfar}
\end{equation}
It varies from 1/3 at the resonance to 1 at large $\epsilon_d^2$. We also compare the asymptotics (\ref{omegaMfar})
with \cite{prb,epl}:
\begin{equation}
\omega^2_I=\omega^2_0 -  \Gamma^2
\frac{4 \Delta ^2 (\Delta ^2+\epsilon_d^2)}{\omega_0 ^4}+\omega _0^2 O\left(\frac{\Gamma^4}
{\omega _0^4}\right)
 \,  . \label{omegaIfar}
\end{equation}
Both frequencies, the minimum location  $\omega_M$ and the oscillation
frequency $\omega_I$ being some transformation of the initial qubit frequency $\omega_0$
by the tunneling produced dissipation
are different and approach one another only asymptotically at small $\Gamma$.

A similar feature of finite frequency dip , but  of a smaller depth, in the current noise spectrum
has been predicted \cite{flindt,brandes}
in transport through a Coulomb blockaded double quantum dot under the condition of
not more than its single electron occupancy. This dip is produced by the
displacement part of the total current, which realizes the Coulomb screening of the dots
in this model due to finite capacitances between the dots and the leads.
This mechanism of screening does not change the tunneling particles dynamics and the particle currents. Hence,
it does not affect the low frequency noise and no its spectrum bifurcation has been
found in this system.

\section{Conclusion}

The quantum fluctuations of the current of
spinless electrons tunneling through an
interacting resonant level of a QD into an empty collector have been
studied in the especially simple, but realistic model, in which
all sudden variations in charge of the QD are effectively screened by
a single tunneling channel of the emitter. Making use of the
exact solution to this model, we have derived a general expression for
the counting statistics
of the charge transfer and found a simple relation between
the two statistics for the
processes of the QD evolution from its stationary and empty states.
This relation
has allowed us to obtain the spectrum of the steady
current shot noise
through calculation of the Fourier transformation of
the time-dependent transient current produced in the process of
the QD empty state evolution.

The oscillating behavior of this current results from emergence of
the qubit built of electron-hole pair at the QD and its coherent dynamics in
the wide range of the model parameters \cite{epl}
and can be used for identification of  the FES observed in
the tunneling current dependence on voltage.
We have demonstrated that the current noise spectrum
can also be used for this purpose.
Indeed, its frequency dependence  normalized by
the mean current is characterized by the
dips whose positions reflect the oscillating
behavior of the time-dependent transient current:
A single zero-frequency minimum in the normalized spectral dependence
occurring for the large collector tunneling rate $\Gamma$
splits with decrease of $\Gamma$  into the two resonant dips located at
finite frequencies $\pm \omega_M$ when either the emitter tunneling
coupling $\Delta$ or the absolute
value of the resonant level energy $|\epsilon_d|$  become large
enough in comparison with the collector tunneling rate $\Gamma$
and either $\Delta^2>5\Gamma^2/8$ or $\epsilon_d^2> 0.29 \Gamma^2$ holds.
Note, these conditions are more restrictive than the ones for the transient
current oscillations to appear, which are  either $\Delta>\Gamma/4$ or
$\epsilon_d^2>\Gamma^2/27$, respectively.

\bigskip

\acknowledgments The work was supported by the Leverhulme Trust
Research Project Grant RPG-2016-044 (V.P.) and
Russian Federation STATE TASK No 007-00220-18-00 (I.L.)


\begin{thebibliography}{0}

\bibitem{1}
\Name{G. D. Mahan}
\REVIEW{Phys. Rev.}  {163}{1967}{612}.
\bibitem{2}
\Name{P. Nozieres \and C. T. de Dominicis}
\REVIEW{Phys. Rev.} {178}{1969} {1097}.
\bibitem{3} \Name{P. H. Citrin}
\REVIEW{Phys. Rev. B} {8}{973}{5545}.
\bibitem{4} \Name{P. H. Citrin, G. K. Wertheim, \and Y. Baer}
 \REVIEW{Phys. Rev. B} {16}{1977}{4256}.
\bibitem{5} \Name{K. A. Matveev \and A. I. Larkin}
\REVIEW{Phys. Rev. B} {46}{1992}{15337}.
\bibitem{geim} \Name{A.~K.~Geim, P.~C.~Main, N.~La~Scala~Jr., L.~Eaves,
T.~J.~Foster, P.~H.~Beton, J.~W.~Sakai, F.~W.~Sheard, M.~Henini., G.~Hill
and M.~A.~Pate} \REVIEW{Phys. Rev. Lett.}{72}{1994} {2061}.
\bibitem{6} \Name{I. Hapke-Wurst, U. Zeitler, H. Frahm, A. G. M. Jansen,
R. J. Haug, \and K. Pierz}
\REVIEW{Phys. Rev. B} {62}{2000}{12621}.2.
\bibitem{7} \Name{H. Frahm, C. von Zobeltitz, N. Maire, \and R. J. Haug}
\REVIEW{Phys. Rev. B} {74}{2006}{ 035329}.
\bibitem{8} \Name{N. Maire, F. Hohls, T. L\"{u}dtke, K. Pierz, \and R. J. Haug}
\REVIEW{Phys. Rev. B }{75}{2007}{233304}.
\bibitem{9} \Name{M.  Ruth,  T.  Slobodskyy,  C.  Gould,  G.  Schmidt,  \and
L. W. Molenkamp}
\REVIEW {Applied Physics Letters}{93}{2008}{182104}.
\bibitem{lar} \Name{I.A. Larkin, E.E. Vdovin, Yu.N. Khanin, S. Ujevic \and M. Henini}
\REVIEW{Phys. Scripta}{82}{2010}{038106} .
\bibitem{prb} \Name{V.V. Ponomarenko,\and I.A. Larkin}
\REVIEW{Phys. Rev. B}{95}{2017}{205416}.
\bibitem{epl}
\Name{V.V. Ponomarenko \and I. A. Larkin} \REVIEW{EPL} {113}{2016}{67004}.
\bibitem{L} \Name{A. J. Leggett}
\REVIEW {Rev. Mod. Phys.} {59}{1987}{1}.
\bibitem{Sch} \Name{O. Katsuba \and H. Schoeller}
\REVIEW {Phys. Rev. B} {87}{2013}{ 201402(R)}.
\bibitem{noise} \Name{N. Ubbelohde {\it et al.}}  \REVIEW{Sci. Rep.}{2}{2012}{374}.
\bibitem{schotte} K. Schotte and U.  Schotte
\REVIEW{Phys. Rev.}{182} {1969}{479} .
\bibitem{lll} \Name{L. S. Levitov,
H.-W. Lee, \and G. B. Lesovik} \REVIEW{J. Math. Phys.}  {37} {1996} {4845}.
\bibitem{GKS} \Name{ V. Gorini, A. Kossakowski, \and E. C. G. Sudarshan}
\REVIEW{J. Math. Phys.} {17}  {1976} {821}.
\bibitem{Lindblad} \Name{ G. Lindblad}
\REVIEW{Commun. Math. Phys.} {48} {1976} {119}.
\bibitem{us} \Name{H.T. Imam, V.V. Ponomarenko,\and D.V. Averin}
 \REVIEW{Phys. Rev.} {50} {1994}{18288}.
\bibitem{flindt} \Name{ C. Flindt {\it et al.}}
\REVIEW{Phys. Rev. Lett.}{100}{2008} {150601}.
\bibitem{brandes} \Name{  R. Aguado and T. Brandes}
\REVIEW{Phys. Rev. Lett.}{92}{2004} {206601}.



\end{thebibliography}
\end{document}